\documentclass[prl,showpacs,preprintnumbers,amsmath,amssymb]{revtex4}
\usepackage{amsmath}
\usepackage{amsfonts}
\usepackage{amssymb}
\usepackage{color}
\usepackage{tikz}
\usepackage{feynmf}
\usepackage{bm}
\usepackage{cleveref}
\usepackage{graphicx}
\usepackage{bm}
\usepackage{ dsfont }
\newcommand{\beq}{\begin{equation}}
\newcommand{\eeq}{\end{equation}}
\newcommand{\beqa}{\begin{eqnarray}}
\newcommand{\eeqa}{\end{eqnarray}}
\newcommand{\bs}{\boldsymbol}

\usepackage{graphicx,amssymb,amsmath,subfigure}
\newcommand{\bb}[1]{\mathbf{#1}}

\newcommand{\mth}[1]{\mathcal{#1}}

\newcommand{\vf}{v_\text{F} }

\begin{document}
\bibliographystyle{naturemag}

\title{Thermoelectric relations in the conformal limit in Dirac and Weyl semimetals }
\author{Vicente Arjona$^{1}$, Juan Borge$^{2}$,  and Mar\'ia A.H. Vozmediano$^{1}$}
\email{${1}$vicente.arjona@hotmail.com, ${2}juanborge@hotmail.com, ${3}$vozmediano@icmm.csic.es}
\affiliation{
${1}$ Instituto de Ciencia de Materiales de Madrid, C/ Sor Juana In\'es de la Cruz 3, 
Cantoblanco, 28049 Madrid, Spain\\
${^2}$ Nano-Bio Spectroscopy group, Departamento de F\`isica de Materiales,
Universidad del Pa\'is Vasco UPV/EHU, E-20018 San Sebasti\'an, Spain}

\begin{abstract}
Dirac and Weyl semimetals are three-dimensional electronic systems with the Fermi level at or near a band crossing. Their low energy quasi-particles are described by  a relativistic  Dirac Hamiltonian with zero effective mass, challenging the standard Fermi liquid  (FL) description of metals.  In FL systems, electrical and thermo--electric transport coefficient are linked by very robust relations. The Mott relation links the thermoelectric and conductivity transport coefficients. In~a previous publication, the thermoelectric coefficient  was found to have an anomalous behavior originating in the quantum breakdown of the conformal anomaly by electromagnetic interactions. We analyze the fate of the Mott relation in the system.  We compute the Hall conductivity of a Dirac metal as a function of the temperature and chemical potential and  show that the  Mott relation is not fulfilled in the conformal limit.   

\end{abstract}

\maketitle

\section{Introduction}
Dirac semimetals are three-dimensional (3D) crystals with band crossings near the Fermi level. In a low energy continuum description   their quasi particles obey the massless Dirac equation and the interacting system is identical to massless quantum electrodynamics (QED). The low energy bands in Dirac semimetals are four fold degenerate (two spins, two chiralities). Breaking inversion or time-reversal symmetry gives rise to Weyl semimetals subjected to many interesting transport phenomena related to the chiral anomaly \cite{XKetal15,Lietal15,ZXetal16}. After their synthesis in 2015, Dirac and Weyl semimetals have evolved to become a main topic in condensed matter physics. In addition to their interest as material realization of high energy phenomena \cite{Karl14}, they present exceptional thermoelectric properties~\cite{GSetal18}.

Leaving aside the important technological applications, thermal and electro--thermal transport are very useful tools
to characterise the physical properties of new materials \cite{BA16}. In standard metals where almost free quasiparticles transport both electric charge and entropy, there are tight relations between the transport coefficients \cite{Sond48,Mott69,Kittel}. The experimental confirmation of these relations are used to characterize the low energy electronic nature of new materials and their possible phase transitions as a function of temperature or doping.  Dirac materials in two (graphene) \cite{XY06,CO09,WB09,PO13,GFetal16} and three \mbox{dimensions \cite{LG13,LLF14,LDS16,GMSS17,MM18,NN18,SK18,MS19}} have been extensively analyzed under the  theoretical and experimental  points of view.  These materials have zero density of states when the Fermi level is at the Dirac point, Coulomb interactions are poorly screened and, in very clean samples, a hydrodynamic regime with breakdown of the thermoelectric relations  has been experimentally established \cite{BTetal16,CJetal16,LDS16,MKetal16,BSetal18,GMetal18,JFetal18}.  

In a recent publication \cite{CCV18}, it was shown that the conformal anomaly \cite{Cherno16}, related to metric deformations,  gives rise to a special contribution to the Nernst signal which remains finite at zero temperature and chemical potential, a very unusual property. This result was later confirmed with a more standard Kubo calculation in \cite{ACV19}. In this work we analyze the validity of the  phenomenological thermoelectric relations in Dirac and Weyl semimetals in the conformal limit (zero temperature and chemical potential). 

We perform an explicit calculation of the Hall conductivity in the conformal limit and, combining it with the thermoelectric coefficient of ref. \cite{ACV19}, we see that the Mott relation, an easy to measure property, is violated in the conformal limit. Since the density of states is zero at the neotrality point we cannot use a Boltzmann formalism, and the calculation is done with a Kubo formula. Extending the calculation to include a finite temperature and chemical potential, we see that the Mott relation is recovered  at finite temperature and chemical potential. This result implies the singularity of the conformal point and the fact that, as it was well known  in graphene,  its physics can not be reached by taking the $\mu\to 0$ limit of a doped system.

\section{Thermoelectric Relations}

Applying  an external electric field ${\bb E}={\bb\nabla} V$ or a temperature gradient ${\bb\nabla} T$ to a conductor,  induces electric ${\bf J}$ and heat currents  ${\bf J}_\epsilon$ that, in linear response, can be written as \cite{LLF14}:
\beq
\begin{pmatrix}
J^i\\
 J^i_\epsilon\\
\end{pmatrix}
=
\begin{bmatrix} \sigma^{ij} & \alpha^{ij}  \\ T\bar\alpha^{ij} & \kappa^{ij} \end{bmatrix}  
\left( \begin{array}{c} E_j \\-\nabla_j T
\end{array} \right),
\label{eq_main}
\eeq
where   $\sigma$, and $\kappa$ are the electric and thermal conductivity tensors respectively,  and $\alpha$ is the thermoelectric coefficient. $i$ and $j$ index spatial dimensions, i.e., $i, j = x,y,z$. 
In the presence of a magnetic field (in the
chosen z-direction along OZ axis, 
the various tensors in \eqref{eq_main} can have time-reversal odd off-diagonal components as the Hall conductivity $\sigma^{xy}$ or the  transverse components of  $\alpha^{xy}$.

The best known phenomenological laws used in thermo--electrical transport are the Wiedemann-Franz (WF) law and the Mott relation \cite{Ziman,JM80}:
\beq 
\sigma_{ij}=T{\mathcal L}\kappa_{ij}, \qquad 
\alpha_{ij}=T{\mathcal L}e\left(\frac{d \sigma_{ij}}{d\mu}\right)_{\mu=\epsilon_F}.
\label{eq_Mott}
\eeq

The first one establishes  that the ratio
of the thermal to the electrical conductivity is the temperature times a universal number,
the Lorenz number ${\cal L}=\pi^2k_B^2/3e^2$, where $k_B$ and $e$ are the Boltzmann constant and the unit charge respectively. The Mott relation relates the  tensor $\hat\alpha$ with the temperature times the value of the derivative of the electrical conductivity with respect to the chemical potential at the Fermi level. In the presence of  magnetic fields magnetization currents can arise. The~Mott relation holds for transport currents only \cite{SS77,JG84,CHR97,QTN11}. The validity of these laws has been established for any system which can be described as a Fermi liquid, provided the quasiparticles do not exchange energy during collisions. It has also been proven to be valid  when a semiclassical description of the electronic system is allowed. Deviations from these phenomenological relations in conventional matter are normally attributed to electron--electron interactions inducing departures from Fermi liquid behavior or to the emergence of a new phase regime \cite{AKS11,BSetal18,GMetal18}. 

An interesting question arose associated to the thermoelectric relations in topological materials. These materials have anomalous conductivities (particularly Hall conductivity) similar to that occurring in ferromagnetic materials induced by the Berry curvature of the bands.  The question of whether or not these anomalous transport coefficients obeyed the WF and Mott relations, arose soon after the recognition of topological properties. The validity of the Mott relation for the anomalous transport phenomena was observed experimentally in films of $Ga_{1-x}Mn_xAs$, a ferromagnetic semiconductor in~\cite{YCetal08}. Berry curvature effects were cleverly added in the Boltzmann transport formalism  in a way to fulfil   the rules \cite{XY06,RML12} but the experimental answers to these question are more diverse \cite{YCetal08,Kim14,LGetal14,Lietal15}. 

Typically Dirac materials in two and three dimensions are expected to follow the standard relations in the low T regime   and deviate from it at larger temperatures \cite{LLetal17,GMSS17,JFetal18}. Violations of the WF law have been described in these systems associated to the presence of a hydrodynamic regime where departures from the standard FL behavior are also found  \cite{LDS16,GMetal18}. The thermoelectric properties of Dirac and Weyl semimetals are a very active field of actual research \cite{LLetal17,WMetal18,SMetal18,SK18}. 
The influence of lattice deformation on the thermoelectric transport properties in Weyl semimetals has been discussed in \cite{MS19}.

\section{Thermoelectric Coefficient in the Conformal Invariant Point in Dirac Semimetals}

In the next section we will analyze the Mott relation at the light of the results in \cite{CCV18,ACV19}  that we will summarize here. Close to  a two bands band crossing in three spacial dimensions, the dispersion relation of an electronic system can be linearized giving rise to the effective low energy  Hamiltonian~\cite{AMV18}
\beq
H_s=s v_F \boldsymbol\sigma \cdot {\bf k}. 
\label{eq:Ham}
\eeq
where $s=\pm$ is the chirality (left, right) of the band, $\sigma$ are the three Pauli matrices, and ${\bf k}$ is the momentum of the quasiparticles. These two 2-dimensional Hamiltonians emerge from a 4-dimensional massless Dirac equation.

The action associated to this Hamiltonian in the presence of a background electromagnetic potential $A_\mu$  is
\beq
S=\exp\{-i\hbar\int L(x) d^4 x\}, \quad L(x)= \bar\Psi \gamma^\mu (\partial_\mu+ie A_\mu)\Psi,
\label{eq:action}
\eeq
where $\bar\Psi\equiv\Psi^+\gamma^0$ and $\gamma^\mu$ are four dimensional Dirac matrices. 
Aside from the fact that the Fermi velocity replaces the speed of light in the space components of the current, this  is the action of massless quantum electrodynamics (QED).
Since the action has not dimension--full parameters, it is invariant, at the classical level,  under a scale transformation: $x\to \lambda x , \Psi\to \lambda^{-3/2}\Psi$, $A_\mu\to \lambda^{-1}A_\mu$. By the Noether theorem there is an associated dilatation current: $d^\mu=x_\nu T^{\mu\nu}$ whose conservation implies the vanishing of the trace of the stress tensor: $T^\mu_\mu=0$. After quantizing the action, the  scale invariance acquires a quantum anomaly proportional to the electromagnetic field strenght 
\cite{CDJ77}.
This anomaly was shown to give rise, in the context of high energy physics,  to the scale magnetic effect,  an electric current perpendicular to an applied magnetic field and a gradient of the conformal factor \cite{Cherno16}.
In \cite{CCV18}, the scale magnetic effect was promoted to a  contribution to the Nernst effect in the material realization of conformal QED provided by Dirac or Weyl semimetals. A non-zero thermoelectric coefficient was predicted at zero temperature and chemical potential where the theory is scale invariant at the classical level. A  Kubo formula calculation of the thermoelectric coefficient extended the calculation to finite temperature and chemical potential and  gave the same result in the conformal limit \cite{ACV19}.

Chosing  the magnetic field to point in the OZ direction ( $B_z$), and the temperature gradient  in the OY direction $\nabla_y T$,  the induced thermoelectric current points in the OX direction. In this geometry, the~expression \eqref{eq_main} reads $$J^x= \alpha^{xy}\nabla_y T.$$

The coefficient $\alpha^{xy}$ for a single Dirac cone at zero chemical potential ad in the limit $\nabla T/T=0$  was obtained to be
\beq
\alpha^{xy}=\frac{e^2 v_F B}{4 \pi^2  \hbar T}  .
\label{eq_chi}
\eeq

\section{The Hall Conductivity}

In order to analyze the Mott relation (second equation in \eqref{eq_Mott}), we need to compute the derivative of the Hall conductivity as a function of the chemical potential. 
The appropriate  expression \eqref{eq_main} reads~now $$J^x= \sigma^{xy} E_y .$$

In most works on topological metals,  the Hall conductivity is calculated using a Boltzmann formalism for the electronic transport. Since in Dirac semimetals the density of states at the neutrality point is zero, a Boltzmann approach does not seem reliable. 

In this work 
the Hall conductivity is computed with a Kubo formulation as the one done in~\cite{ACV19} for the thermoelectric coefficient. 
In the Kubo approach, the conductivity tensor is given by the expression: 
\beq 
\sigma^{ij} (\omega, \bb q ) = - \frac{1}{\omega \mth V } (2\pi)^3 \int \! \text{d}t' \, \frac{1}{\hbar}\Theta (t-t')\cdot 
\left \langle \left [ J^i (t, \bb q  ) , J^j (t', - \bb q) \right] \right \rangle,
\label{eqn:sigmamom}
\eeq 
where $\mth V $ is the volume of the system and the current operator associated to the action \eqref{eq:action} is given by
\beq
J^i=v_F\bar\Psi\gamma^i\Psi,
\eeq
where ($i=x,y,z$).
We will compute the Hall component of the conductivity tensor $\sigma^{xy} (\omega, \bb q ) $ for  a Weyl semimetal in an external  magnetic field in the $0Z$ direction. 

The magnetic field is coupled to  the Hamiltonian 
\eqref{eq:Ham} by a Peierls substitution ${\bf k}\to({\bf k+A})$. 
In the Landau gauge $A_x= -B y$, the spectrum of the system is:  
\beq 
E_n(k_z)  = \text{sign}(n) \vf \left[ 2 e \hbar B \vert n \vert  + \hbar^2 k_z^2 \right]^{1/2}      \hspace{1cm}  n\in \mathbb{Z}, \; n\neq 0\;,
\eeq 
\beq 
E_{0 s}(k_z) =  s \vf \hbar k_z ,
\label{eq_zeroLL}
\eeq 

\begin{figure}
\centering
\includegraphics[width=0.6\columnwidth]{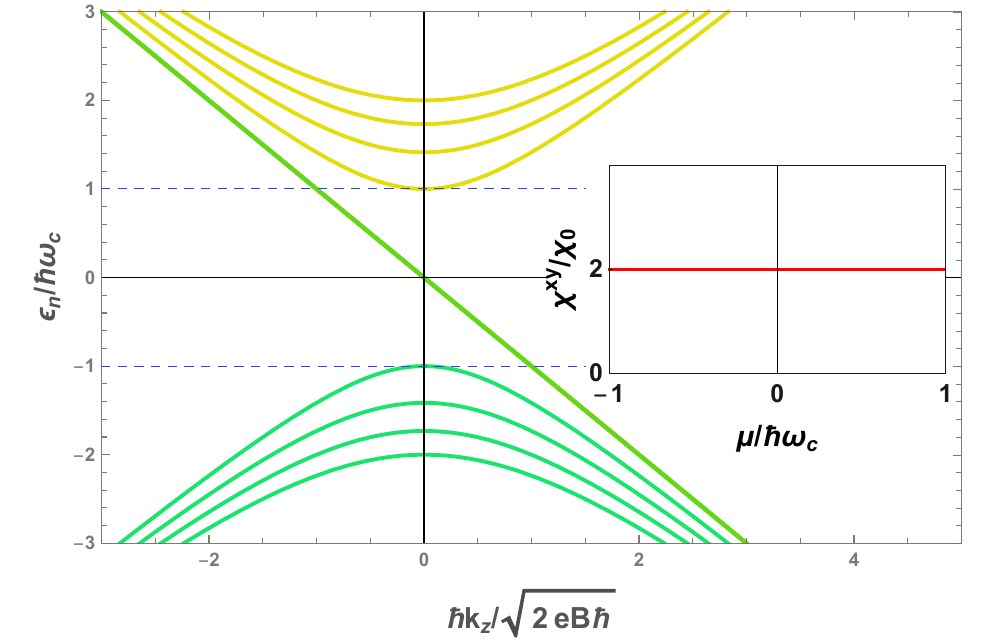}
\caption{Landau level structure of a single chirality in a Dirac semi-metal. The green straight line represents the chiral zeroth LL. The inset shows the thermoelectric coefficient $\chi^{xy}=\alpha^{xy}/T$ computed in ref. \cite{ACV19}.}
\label{fig:xit}
\end{figure}

The presence of the zeroth Landau level (LL) \eqref{eq_zeroLL} is one of the most prominent characteristics of Dirac matter in applied magnetic field. This  one-dimensional chiral fermion is ultimately responsible for the chiral anomaly and the anomalous Hall conductivity in the materials \cite{NN83}. It has a linear dispersion and a constant density of states at fixed magnetic field. Figure \ref{fig:xit} shows the Landau level (LL) structure of a single chirality in a Dirac or Weyl semimetal.  

The green straight line represents the chiral zeroth order LL. The inset shows the zero temperature thermoelectric coefficient $\chi=\alpha/T$ normalized to the value
$\chi_0 = \vf e^2 B / 4(2\pi)^2 \hbar$ as a function of the chemical potential  computed in ref. \cite{ACV19}.  The function has a constant value  when $\mu$ lies in the interval between the first Landau levels $n=\pm 1$. We will compute the Hall conductivity as a function of the chemical potential in the same conditions as described here.

The Landau eigenfunctions are:

\beq 
\varphi_{\bb k n s } (\bb r) = \frac{1}{\sqrt{L_x L_z }} \frac{e^{i k_x x} e^{i k_z z}}{\sqrt{\alpha_{k_z n s}^2+1}}e^{-(y-k_x l_B^2)^2/2l_B^2} \begin{pmatrix} 
 \frac{\alpha_{k_z n s}}{\sqrt{2^{N-1} (N -1)! \pi^{1/2} l_B } } H_{N - 1} \left[\frac{y-k_x l_B^2 }{l_B}\right] \\ \frac{1}{\sqrt{2^{N } N ! \pi^{1/2} l_B } } H_{N } \left[\frac{y-k_x l_B^2 }{l_B}\right]
\end{pmatrix} ,
\label{eqn:lwe}
\eeq
with
\beq
\alpha_{k_z n s} =\frac{-\sqrt{2 e B \hbar \vert n \vert }}{E_{k_z n s}/s \vf - \hbar k_z} .
\eeq 

Capital letters refer to the absolute value of Landau levels, $H_n (x)$ are Hermite polynomials, and the factor $\sqrt{\alpha_{k_z n s}^2 +1 } $ comes from the wave-function normalization.

In the Landau level basis and using the Lehman representation of the Green's function, the Hall conductivity in the local and zero frequency limit is given by the expression: 
\beq
\sigma^{xy}= \lim_{\eta\to0} \sum_{ m,n} 
\frac{1}{4 \sqrt{2} \pi^2}\frac{ e^2}{\hbar l_B}
\int \! \text{d}\kappa_z \frac{- 2 \alpha_{ \kappa_z m s }^2 }{ (\alpha^2_{\kappa_z m s } + 1 ) ( \alpha^2_{\kappa_z n s } + 1 )} \frac{ n_{\bs \kappa m s } - n_{\bs \kappa n s} }{(\epsilon_{\kappa_z m s } -\epsilon_{\kappa_z n s} + i \hbar \eta  )^2 } \;.
\label{eqn:sigma00}
\eeq

In \eqref{eqn:sigma00}  the sum runs over the Landau level index  and is restricted to $\vert m\vert= \vert  n \vert\pm 1$. $l_B$ is the cyclotron length, and $n_{\bs \kappa m s }$ is the Fermi--Dirac distribution function which depends on the Landau level $m$, and the chirality $s$ through the dispersion relation. The integral  is written in terms of the dimensionless variables $\sqrt{2eB\hbar}\kappa_z = \hbar k_z$, $ \epsilon_{\kappa_z m s} = E_{k_z m s} / \hbar \omega_c $ ($\omega_c$ is the cyclotron frequency) and the factor $e^{-\eta t}$ has been introduced to guarantee the convergence of the time integral.

\begin{figure}
\centering
\includegraphics[width=0.7\columnwidth]{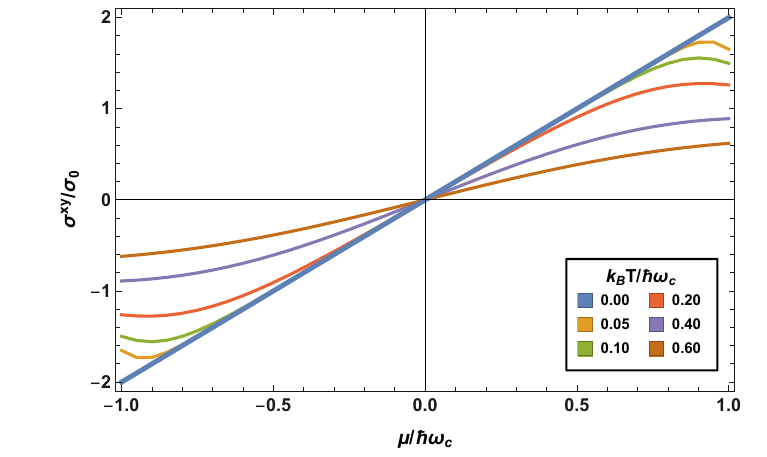}
\caption{The Hall conductivity $\sigma^{xy}$ computed in this work as a function of the chemical potential in the range  $-\hbar \omega_c \le \mu \le \hbar \omega_c $ for various values of the temperature.  {At $T=0$, the conductivity  is linear in $\mu$.} }
\label{fig:sigmat}
\end{figure}

For completeness we have also performed the calculation at finite chemical potential and temperature. These variables enter the Kubo expression \eqref{eqn:sigma00} through the Fermi--Dirac distribution function. Our results are summarized in Figure  \ref{fig:sigmat}.

Figure  \ref{fig:sigmat} shows the the Hall conductivity $\sigma^{xy}$ as a function of the chemical potential for the interval $-\hbar \omega_c \le \mu \le \hbar \omega_c $ and  for various values of the temperature.  The chosen interval is the same used in Figure  \ref{fig:xit} and corresponds to the quantum limit where only the zeroth Landau level is filled.
The conductivity is normalized to the value $\sigma_0 = \frac{1}{\sqrt{2} 4 \pi^2}\frac{s^2 e^2}{\hbar l_B} $.
For $T=0$, the conductivity is linear in $\mu$.  
As the temperature increases, the slope decreases and the function becomes smoother due to the contribution of the thermally activated carriers at higher Landau levels.
The behavior of the  conductivity is smooth at the neutrality point $\mu=0$.

In the next section we will analyze the Mott relation.
\section{The Mott Relation}
The Mott relation \eqref{eq_Mott} can be written as
\beq
(\alpha/\partial_\mu\sigma)_{\mu=E_F}={\cal L}T.
\eeq
where ${\cal L}$ is the Lorenz number appearing in the Wiedemann-Franz law. 
Given the  finite value of the thermoelectric coefficient $\chi$ at $T=0, \mu=0$ and with our result on the smooth behavior of the conductivity at this point, it is clear that the relation does not hold at the conformal point. 
 
We have analyzed the situation at finite chemical potential and finite temperature performing a numerical calculation of the expression \eqref{eqn:sigma00}. 
 
Inserting the thermoelectric conductivity $\alpha^{xy}$ in \eqref{eq_Mott} we get: 
\beq 
\chi^{xy} \left( \frac{\partial \sigma^{xy}}{\partial \mu} \right)^{-1}_{\mu=E_\text{F}} = \mth R  T^2.
\label{eqn:mott1}
\eeq 
 
\begin{figure}
  \centering
\includegraphics[width=0.75\columnwidth]{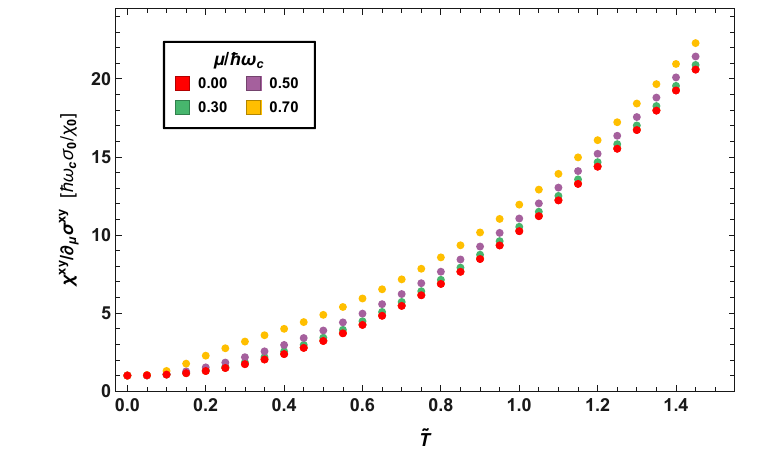}
\caption{Behaviour of $\chi^{ij}/\partial_\mu \sigma^{ij}$ as a function of the temperature. The Mott relation is not satisfied for $T=0$, where the ratio presents a finite value.}
\label{fig:T}
\end{figure}

The ratio  $\chi^{ij}/\partial_\mu \sigma^{ij} $ as a function of the dimensionless temperature variable $\tilde T = k_B T / \hbar \omega_c $ is plotted in Figure  \ref{fig:T} for different values of the chemical potential. 
We notice that at zero temperature and for $\mu=0$ (red points) the relation is not satisfied, the function presenting a finite value at $T=0$. As the temperature increases, it follows a quadratic behaviour for all values of $\mu$. 

\begin{figure}
\centering
\includegraphics[width=0.6\columnwidth]{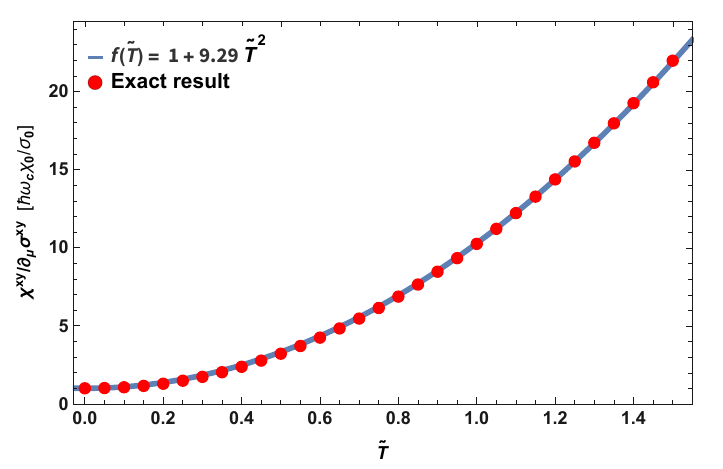}
\caption{Temperature dependence of the Mott ratio between the thermoelectric response function $\chi^{xy}$ and the derivative of the electric conductivity $\partial_\mu \sigma^{xy}$ at $\mu=0$. Red dots are the numerical calculation and the blue line is the fit to the function $f(\tilde{T}) =  1 + 9.27 \tilde{T}^2$. The deviation is smaller than the size of the points.}
\label{fig:mott}
\end{figure}

Figure  \ref{fig:mott}  shows a fit of the numerical values computed at various temperatures (exact results) to a continuum function  providing the expression $f(\tilde{T}) =  1 + 9.29 \tilde{T}^2$,
where is the dimensionless temperature variable defined above: $\tilde T = k_B T / \hbar \omega_c $. 

Restoring the units we get, away from $T=0$, the coefficient of the parabola  ${\cal R}= 2.32 (k_B)^2/e$, which coincides with the standard value of 
the Lorenz number ${\cal L}= 2.44 (k_B)^2/e$ \cite{KPP93}
to a great accuracy.

\section{Discussion}

The main conclusion of this work is the violation of the Mott relation at the conformal point of Dirac matter. In particular, a previous calculation \cite{CCV18,ACV19} showed an unexpected finite term in the thermoelectric coefficient in this limit. In order to fulfill the Mott relation at the conformal point, the~Hall conductivity should have developped a singular behavior at $T=0$, $\mu=0$. Our explicit Kubo calculation \eqref{eqn:sigma00} shows a smooth behavior of the conductivity in that limit, implying that charge and heat transport are not related by the standard quasiparticle description in the neutral system. 

Our results show that, as also happens in graphene, the $\mu=0$ point in clean Dirac semimetals is a singular point (the physical properties of the system cannot  be obtained as the limit $\mu\to 0$ of the finite $\mu$ system) and  away from it, the materials show standard Fermi liquid behaviour.  It is also interesting to note that lattice effects occurring at higher energies \cite{GMSS17}  do not alter significantly the general behavior of the thermoelectric coefficients. 
Our work points, in general, to  the singular behavior of Dirac materials when the Fermi energy lies close to the Dirac point. This is where the correspondence with the high energy counterparts works better.  In our case, the non zero thermoelectric coefficient found
in~\cite{ACV19}
and the consequent breakdown of the Mott relation at the neutrality point analyzed in this work
has its root in the conformal anomaly described in quantum field theory. 
It is interesting to note that the conformal anomaly is a part of the gravitational anomalies whose influence in the thermal transport is a very hot subject in todays condensed matter experiments \cite{Getal17,SGetal20,VZetal19}.
The analysis  of the present work deepens the relation between the high and low energy phenomena manifested in these~materials.

\begin{acknowledgments}
We thank B. Bradlyn, A. Cortijo and  Y. Ferreiros, for interesting discussions.
This work  has been supported by the PIC2016FR6/PICS07480, Spanish MECD grant FIS2014-57432-P, European Union structural funds and the Comunidad Aut\'onoma de Madrid (CAM) NMAT2D-CM Program (S2018-NMT-4511). J. B. acknowledges the financial support from the European Research Council (ERC-2015-AdG-694097) and Spanish Ministry (MINECO) Grant No. FIS2016-79464-P.
\end{acknowledgments}


\bibliography{Nernst}

\end{document}